\documentstyle[11pt,twoside,pasp3D,epsf]{article}

\begin{document}

\title{Infrared 3D Observations of Nearby Active Galaxies}

\author{R. Maiolino$^1$, N. Thatte$^2$,
A. Alonso-Herrero$^3$, D. Lutz$^2$, A. Marconi$^1$}

\affil{$^1$Osservatorio Astrofisico di Arcetri, Firenze, Italy\\
$^2$Max-Planck-Institut f\"{u}r Extraterrestrische Physik, Garching, Germany\\
$^3$Steward Observatory, Tucson, Arizona}

\begin{abstract}
We present multi-wavelength imaging observations of three nearby and
famous active galaxies obtained with NICMOS, ISOCAM and the MPE
near-IR integral field spectrometer. The data reveal a variety of features and
properties that are missed in optical studies and in traditional
IR monodimensional spectroscopy.
\end{abstract}

\section{Introduction}

Infrared observation of active galaxies (starbursts and AGNs) have greatly
improved our understanding of these systems, not only because they are
usually heavily obscured at optical wavelengths, but also because the infrared
bands offer a wealth of indicators
that often do not have an equivalent in the
optical range.
Over the past few years, new IR instruments and observing facilities
have provided two-dimensional spectroscopic information of these
systems through integral field spectroscopy and narrow band imaging.
New integral field spectra obtained with the MPE 3D near-IR spectrometer,
NICMOS-HST narrow and broad band images and ISOCAM-CVF spectra of three
nearby and famous active galaxies allowed a detailed investigation of their
nuclear region revealing new features and properties, and showing that
optical observations and traditional IR spectroscopy only provide
a limited view of this class of objects.
Here we summarize some of the results based on these data.

\section{NGC1068}

This famous galaxy is regarded as the archetype of Seyfert 2 galaxies.
The [OIII]5007\AA \ map shows an ionization cone to the North-East
 of the nucleus, and a much fainter counter-cone to the South-West.
The morphology and the dynamics of the Narrow Line Region (NLR) appear
affected by the radio jets (eg. Axon et al. 1998).
The radio jets also feed two large scale radio lobes.

We observed the nuclear region of NGC1068 with 3D, the MPE near-IR integral
field spectrometer (Weitzel et al. 1996), assisted by ROGUE, a first order
adaptive optics system (Thatte et al. 1995), at the William Herschel Telescope
(WHT)
and at the Anglo Australian Telescope (AAT).

\begin{figure}[!ht]
\plotfiddle{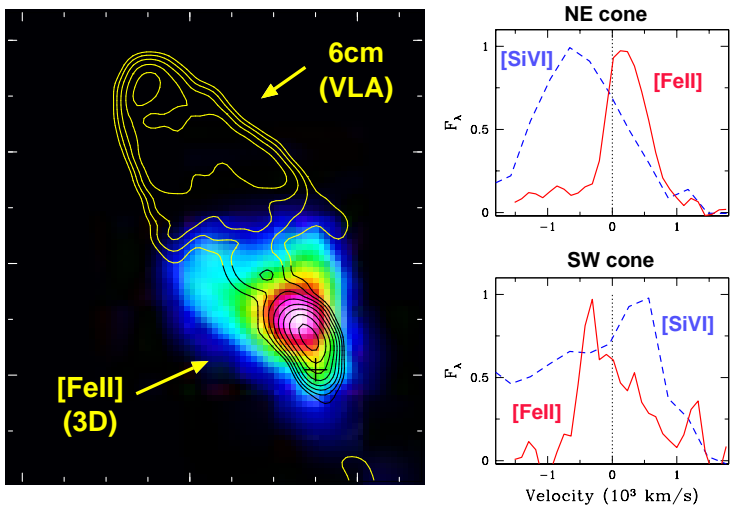}{2.5in}{0}{55}{55}{-150}{-190}
\caption{NGC1068. Left:
itegrated [FeII]1.64$\mu$m emission obtained with the MPE-3D
integral field spectrometer overlayed on the 6cm VLA map of the NE lobe.
Right: [FeII]1.64$\mu$m and [SiVI]1.97$\mu$m line profiles in the two
ionization cones.
}
\end{figure}

Various IR lines and powerful
hot dust emission were detected (Thatte
et al. 1997). Here we focus on the properties of the
[SiVI]1.96$\mu$m high excitation coronal line and the [FeII]1.64$\mu$m line
that instead traces the low ionization region. Both lines are detected in the
NE ionization cone and, at a lower level, in the SW cone. Overall, the
morphology
of their integrated emission is similar to the [OIII]-HST map, though at a
lower resolution. In Fig.~1 we show the [FeII] map, which has the
characteristic cone-like morphology. Given that both these lines are observed
in the ionization cones it is tempting to assume that they are emitted by the
same population of ionized clouds.
However, the kinematics of the gas as traced by the two lines points
to a more complex scenario. In Fig.~1 we show the profile of the [FeII]
and [SiVI] lines, with respect to the host galaxy systemic velocity.
In the NE cone [SiVI] is blueshifted while [FeII] is
redshifted and the shift between the two lines is several 100 km/s. In the SW
cone the situation is exactly reversed. The large velocity difference between
the two lines indicates that they come from different populations of ionized
clouds.

We think that the radio lobes affect the kinematics and the ionization
structure of the large scale NLR, similar to that observed on smaller
scale clouds near the radio jet (Axon et al. 1998). Fig.~2 schematically
illustrates our model. In the NE radio lobe (right-hand side in
Fig.~2) the upper part of the expanding bow shock accelerates the gas in our
direction, hence its emission is blueshifted. This gas is mostly out of the
galactic gas disk and, therefore, it has low density, hence high ionization
parameter that favors the emission of high ionization species such as Si$^{+6}$.
On the opposite side the bow shock enters the dense gas of the galactic disk,
that is therefore redshifted. Here the bow shock rapidly loses its energy, a
fraction of which destroys dust grains returning Fe into the gas phase, thus
increasing the [FeII] emission. Also, the high density
characterizing this region decreases the ionization parameter, while the
column of gas and dust in these equatorial regions hardens the ionizing photon
flux; both these effects favor the emission of low ionization lines typical
of the transition regions, such as [FeII]. The symmetric model for the SW
cone explains the velocities of [FeII] and [SiVI] inverted with respect to the
NE cone.

\begin{figure}[!ht]
\plotfiddle{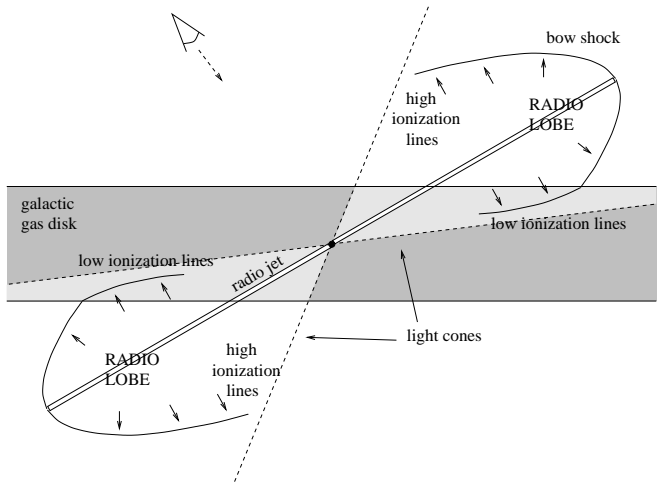}{2.2in}{-90}{40}{40}{-180}{200}
\caption{Proposed model for the Narrow Line Region of NGC1068
}
\end{figure}

Although the bow shock
returns Fe into the gas phase and produces the observed redshift of the low
ionization gas, probably it does not play a major role in exciting the line.
Indeed, the [FeII] map has a cone-like morphology, indicating that
the nuclear X-ray source is probably responsible for exciting this line.

Finally,
the comparison of the radio and [FeII] maps in Fig.~1 further supports our
model.
The lower end of the radio lobe in Fig.~1 nicely
matches the upper end of the [FeII] cone, we think that this dividing
line traces the region where the radio lobe enters the galactic gas disk.
Therefore, the dense gas south of this line has been Fe-enriched and
redshifted, as actually observed. A more detailed discussion of these
results is given in
Thatte et al. (in prep.).

\section{Circinus}

At a distance of only 4 Mpc,
the Circinus galaxy is the closest {\it bona fide} Seyfert 2 known.
Its proximity makes this galaxy an excellent candidate
with which to tackle various issues related to the
interaction between active nuclei and their circumnuclear region.
Within this context, the mechanism responsible for feeding AGNs on the
10--100 pc scale is a debated issue.
Shlosman et al. (1989) and Wada \& Habe (1992) proposed that when the mass
of the nuclear gas disk is a significant fraction of the dynamical mass
($>$20\%) it becomes gravitationally unstable and forms a gaseous bar that
can drive gas into the innermost region to fuel the AGN. Circinus has an
extremely gas-rich nuclear region and, therefore, it is an optimal candidate
to search for this nuclear gaseous bar.

Broad band and narrow band NICMOS-HST
observations of the nuclear region of Circinus were obtained. Fig.~3 shows the
H--K color map of the nuclear region (Maiolino et al. 1999). The galaxy
major axis is at P.A.$\approx 25^{\circ}$ and the SE is the near side of the
disk.
The PSF of the nucleus, dominated by hot dust emission (Maiolino
et al. 1998), has been subtracted in Fig.~3.
 In the circumnuclear region the H--K map
traces the effect of dust reddening. The most intersting feature of this map
is the L-shaped dusty feature located to the South-East of the nucleus.
We identify the radially extended part of this feature
with the gas bar expected to feed the AGN in this system according to
models. The deprojected length of the gas bar is about 100 pc. The gaseous
nature of this bar is inferred by the lack of a stellar counterpart even in
the K band light, where extinction is greatly reduced.

Our interpretation can be checked by looking at the kinematics of the
molecular gas. Shocks and torques on the leading side of the bar should
remove angular momentum from the gas which should result in a strong
velocity gradient and radial inflow motions in this region. We observed the
nuclear region of Circinus with 3D+ROGUE at the AAT.
Fig.~3 shows the velocity field of the H$_2$ line at 2.12$\mu$m. The
velocity field of the molecular gas along the leading side of the bar is
characterized by a strong velocity gradient and a highly redshifted component
(i.e. radially inflowing), just as expected by models. This map not only
supports our interpretation of the gas bar, but directly shows radial inflow
motions that are probably responsible for the AGN fuelling.

\begin{figure}[!ht]
\plotfiddle{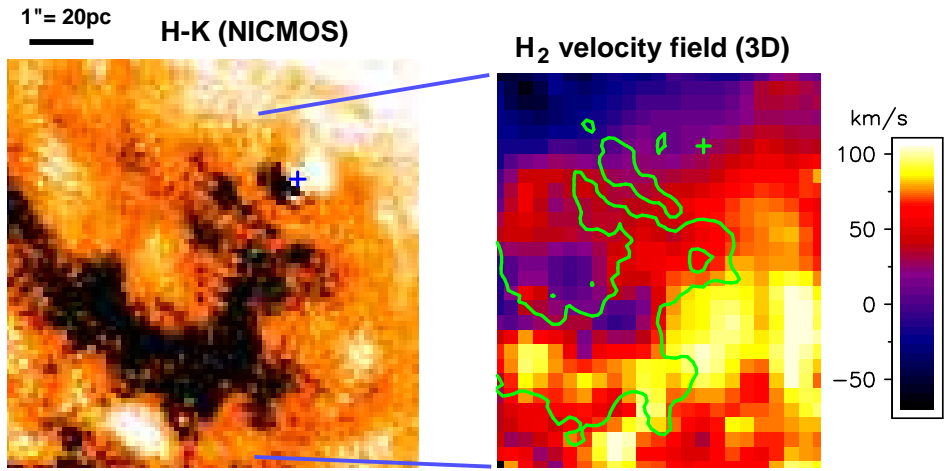}{2.25in}{0}{65}{65}{-190}{-300}
\caption{Circinus. Left: H--K NICMOS map; redder colors corresponf to 
darker regions.  Right velocity field of the
H$_2$2.12$\mu$m line obtained with the MPE-3D spectrometer. The cross marks
the K-band nucleus.
}
\end{figure}

The large amount of molecular gas driven into the central region
also triggers nuclear star formation. Indeed, both
Pa$\alpha$ and [FeII] narrow band NICMOS images reveal diffuse
circumnuclear emission out of the ionization cone that very likely traces
recent star forming activity. A young nuclear stellar population is also
inferred from the low mass-to-light ratio as derived from the CO stellar
bands at 2.29$\mu$m observed in the 3D data (Maiolino et al. 1998).

\section{NGC4945}

Within the context of the starburst-AGN connection, NGC4945
is one of the most spectacular examples of a system where both phenomena
coexist. In the optical and in the IR this edge-on, nearby (3.6 Mpc)
galaxy appears as a heavily obscured starburst.
 However, this galaxy is one of the brightest
extragalactic sources at 100 keV, thus revealing the presence of an active
nucleus that is heavily absorbed along our line of sight ($\rm N_H \simeq
5\times 10^{24} cm^{-2}$).

\begin{figure}[!ht]
\plotfiddle{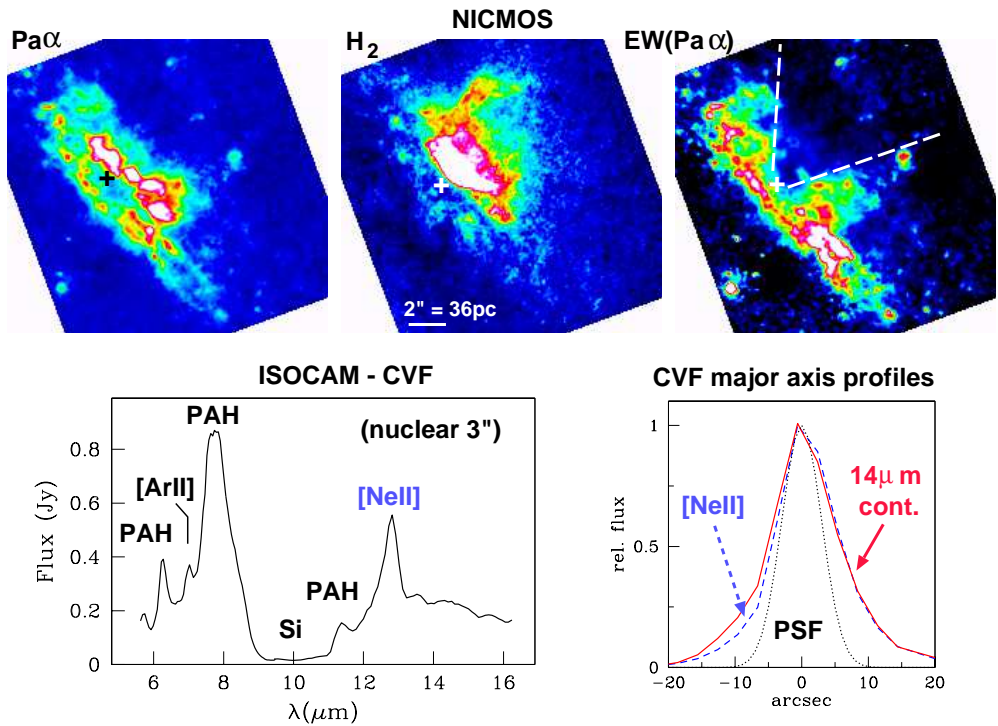}{3.5in}{0}{50}{50}{-205}{-140}
\caption{NGC4945. Top: NICMOS images of
Pa$\alpha$ (1.87$\mu$m), H$_2$ (2.12$\mu$m) and
Pa$\alpha$ equivalent width; the cross marks the location of the H$_2$O
maser.
Bottom: ISOCAM-CVF spectrum of the nuclear
region and profiles of the [NeII]12.8$\mu$m line and 14$\mu$m continuum
emission along the major axis.
}
\end{figure}
NGC4945 was observed with NICMOS both with broad and narrow band filters
(Marconi et al., in prep.).
The Pa$\alpha$ image in Fig.~4 reveals a nuclear starburst
 probably distributed in a ring (about 100 pc in radius)
on the galactic plane. Note that this nuclear region is completely
obscured at optical wavelengths. The K-band image (not shown)
shows no point like source close to the location of the H$_2$O maser
identified by Greenhill et al. (1997). Also,
the CO-index map (derived from the
NIC2-F237M image) does not provide evidence for a significant dilution of
the CO stellar
features due to emission of hot dust heated by the AGN. Such emission
is instead is clearly seen in other similarly obscured AGNs (eg. NGC1068
and Circinus). This indicates that the AGN is obscured along our line of sight
even at 2.3$\mu$m. The H$_2$ 2.12$\mu$m emission line is
mostly distributed above the galactic plane (Fig.~4) in a cavity created by
the starburst superwind and observed also in the J band images
(Moorwood et al. 1996). The lower gas density in the cavity also results in
a lower Pa$\alpha$ emission with respect to the continuum, hence lower
Pa$\alpha$ equivalent width. Interestingly,
the cavity traced by the EW(Pa$\alpha$) has a cone-like
morphology (Fig.~4), similar to the morphology of the NLR in
several Sy2 galaxies. However, no UV-ionizing photons from the AGN reach the
gas in the cavity. Indeed, even though plenty of gas is present in
the outer parts of the cavity,
as traced by the H$_2$ line and dust filaments, this gas
is little ionized and with a spectrum typical of LINERs (possibly excited
by the superwind shock). Therefore, the AGN radiation must be absorbed even
in the direction of the cone-like cavity. Very likely, the AGN is in a
very early stage, still enshrouded in a 4$\pi$ dusty shell a few pc in size.
Given the huge radiation pressure in this region, the dusty shell cannot
survive very long. When the dusty shell will break out,
then the AGN UV radiation
will photoionize the gas within and above the cone-like cavity,
transforming the latter into an ordinary ionization cone.

Although the active nucleus is completely obscured even in the K band, we might
see its dust emission in the mid-IR, both because the obscuration is further
reduced and because warm dust typically emitting at these wavelengths
($\sim 100$ K) is located at larger distances. For this
reason ISOCAM-CVF observations of NGC4945 were obtained. Fig.~4 shows the
CVF spectrum of the nuclear 3$''$.
This spectrum shows typical starburst features (strong PAH.
[NeII]12.8$\mu$m and [ArII]7$\mu$m emission), but the saturated
silicate absorption feature at 10$\mu$m indicates that the mid-IR
emission comes from a heavily obscured region, with $\rm A_V > 50$ mag. There
is some hot dust continuum emission at 14--16$\mu$m.
However, as shown in Fig.~4, this continuum emission is
resolved along the galaxy major axis and its distribution is similar
to that of [NeII], therefore also the
continuum emission is mostly due to the starburst activity. The active
nucleus is therefore hidden and undetected even in the mid-IR. Higher angular
resolution mid-IR observations are required to check if a weak point-like
source might be present on the nucleus.

\acknowledgments
These results were obtained within a wide collaboration that includes also
S. Anders, R. Genzel, D. Macchetto,
A. Moorwood, E. Oliva, A. Quillen, M. Rieke, G. Rieke, E. Schreier,
E. Sturm, L. Tacconi-Garman and D. Tran.

\end{document}